\documentclass[pra,twocolumn,showpacs]{revtex4}
\usepackage{graphicx}
\usepackage{hyperref}
\usepackage{amssymb,amsmath}
\usepackage{epstopdf}
\begin{document}

\title{Characterizing Quantum Properties of a Measurement Apparatus :\\Insights from the Retrodictive Approach}

\author{Taoufik Amri}
\email{amri.taoufik@gmail.com}
\author{Julien Laurat}
\author{Claude Fabre}

\affiliation{Laboratoire Kastler Brossel,
Universit\'{e} Pierre et Marie Curie, Ecole Normale Sup\'{e}rieure,
CNRS, Case 74, 4 place Jussieu, 75252 Paris Cedex 05, France}

\date{\today}

\begin{abstract}
Using the retrodictive approach of quantum physics, we show that the state retrodicted from the response of a measurement apparatus is a convenient tool to fully characterize its quantum properties. We translate in terms of this state some interesting aspects of the quantum behavior of a detector, such as the non-classicality or the non-gaussian character of its measurements. We also introduce estimators - the projectivity, the ideality, the fidelity or the detectivity of measurements perfomed by the  apparatus -  which directly follow from the retrodictive approach. Beyond their fundamental significance for describing general quantum measurements, these properties are crucial in several protocols, in particular in the conditional preparation of non-classical states of light or in measurement-driven quantum information processing. 
\end{abstract}


\maketitle

\textit{Introduction.$\--$}In the quantum world, the measurement process plays a central role, as it leads to an unavoidable and strong modification of the system which is measured. This singular feature has important consequences concerning the foundations of quantum theory \cite{Neumann1955, Zurek1982}, but it also has many practical implications, because the information obtained through a measurement is often used for driving quantum information processing \cite{KLM,OBrien} or preparing a target state conditioned on the result of this measurement \cite{mandel,Ourjoumtsev2007}. To master as much as possible these conditionings, it is therefore very important to characterize as precisely as possible the quantum properties of the measurement apparatus that one uses.

The first experimental quantum characterization of a detector has been achieved only very recently in quantum optics \cite{QDT1}. By using the quantum detector tomography technique (QDT), it is possible to realize the reconstruction of positive operator valued measures (POVMs) \cite{Helstrom} characterizing any measuring device. This technique, by probing the behavior of its responses with a set of known states \cite{QDT2},  gives a complete characterization of the detector seen as a black-box, i.e. only characterized by its responses and without any assumptions about its internal operation. QDT has opened the path of experimental study of novel concepts such as the non-classicality of detectors, with fundamental significance and particular relevance for experimental quantum protocols based on measurements.

The aim of this paper is to show that the retrodictive approach of quantum physics \cite{Aharonov, Barnett0}, which is complementary to the usual predictive one, provides interesting physical insights on the behaviour of a measurement apparatus in terms of a quantum state: the \textit{retrodicted state}. The quantum properties of a measurement performed by a detector can then be associated with the properties of this state. 
We will first introduce the retrodicted state from mathematical foundations of quantum physics. We will then examine several properties of a measurement apparatus from this perspective. In particular, we will provide estimators of certain properties, such as the ideality/projectivity of measurements or their fidelity with projective measurements. We will finally introduce a precise meaning for the non-classicality of measurements by illustrating its relevance for conditional preparation of non-classical states of light. 

\textit{States and Propositions.$\--$}As a preliminary step, we recall a mathematical result demonstrated in \cite{Busch}, which is the recent generalization of the Gleason's theorem \cite{Gleason1957}. From very general requirements about probabilities and the mathematical structure of the Hilbert space, it provides the general expression of probabilities of checking any proposition about the system. 
First we remind that a proposition $P_{n}$  is a property of the system corresponding to \textit{a precise value for a given observable}. This one is represented in the Hilbert space by \textit{a proposition operator} $\hat{P}_{n}$, which is in the simplest case a projector on the eigenstate corresponding to such a value. In the most general case, it can be represented only by \textit{a hermitian and positive operator}. 

One assumes that the probability $\mathrm{Pr}\left(n\right)$ of checking the proposition $P_n$ on the physical system satisfies the three following conditions:
\begin{enumerate}
\item $0 \leq\mathrm{Pr}\left(n\right)\leq 1$ for any proposition $P_{n}$.
\vspace{-0.1cm}
\item $\sum_{n}\,\mathrm{Pr}\left(n\right)=1$ for any \textit{exhaustive} set of propostions such that $\sum_{n}\,\hat{P}_{n}=\hat{1}$.
\vspace{-0.1cm}
\item $\mathrm{Pr}\left(n_{1}\,\textrm{or}\,n_{2}\,\textrm{or}\,...\right)=\mathrm{Pr}\left(n_{1}\right)+\mathrm{Pr}\left(n_{2}\right)+...$ for any \textit{non-exhaustive} set of propositions such that $\hat{P}_{n_{1}}+\hat{P}_{n_{2}}+...\leq \hat{1}.$
\end{enumerate}
According to this theorem for a system needing predictions (i.e. with a Hilbert space of dimension $D\geq 2$), this probability is given by $\mathrm{Pr}\left(n\right)=\mathrm{Tr}\lbrace\hat{\rho}\,\hat{P}_{n}\rbrace$ in which $\hat{\rho}$ is a hermitian, positive, and normalized operator. This operator allows us to make predictions about any properties of the system, and thus constitutes the most general form of its \textit{quantum state}.

\begin{figure}[t]
\includegraphics[width=0.95\columnwidth]{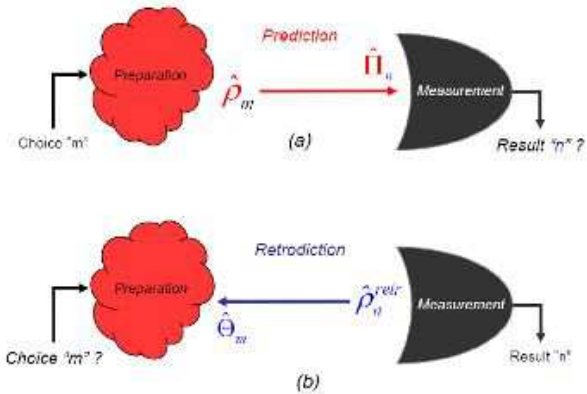} 
\vspace{-1cm}

\caption{(color online)  (a) \textit{Predictive approach} - From a preparation choice \textit{m}, the prepared state $\hat{\rho}_{m}$ resulting from this choice is used for making predictions about subsequent measurements performed on this preparation. (b) \textit{Retrodictive approach} - From a measurement result \textit{n}, the retrodicted state $\hat{\rho}_{ n}^{\mathrm{retr}}$ assigned to the measured system allows retrodictions about state preparations leading to the result \textit{n}.}\label{fig1}
\end{figure}

In quantum physics, any protocol is based on preparations, evolutions and measurements. One can make predictions about the measurement results but also about preparation choices. This corresponds to two approaches: the conventional predictive one, which provides predictions about measurement results, and the less usual retrodictive one, which provides predictions about preparation choices (called retrodictions). According to the previous section, each approach requires a quantum state and an exhaustive set of propositions about it, as pictured on Fig. \ref{fig1}.

In the usual predictive approach,  a preparation of the system in a state $\hat{\rho}_m$ can be associated with a piece of information that we call the \textit{choice} \textit{m}. The prepared state $\hat{\rho}_{m}$ resulting from this choice enables to make predictions about measurement results \textit{n}, labeling the proposition operators which are the POVMs describing the behavior of the responses of the apparatus performing their tests.
From the previous theorem, the probability of obtaining the result \textit{n} after the system was prepared in the state $\hat{\rho}_m$ takes the form:
\begin{equation}
\label{conditional_proba}
\mathrm{Pr}\left(n\vert m \right)=
\mathrm{Tr}\lbrace\hat{\rho}_m \hat{\Pi}_{n}\rbrace,
\end{equation}
where  $\hat{\Pi}_{n}$ is the POVM element corresponding to the result $n$. This is in fact the Born's rule on which the conventional interpretation of quantum physics is based. We now turn to the retrodictive approach and show that the generalization of the Gleason's theorem provides the expression of the retrodicted probabilities, from which the retrodicted state can be derived.

\textit{Retrodictive approach.$\--$}
This approach \cite{Aharonov,Barnett0} is less popular in quantum physics. It has sometimes been used in quantum optics for instance to simplify the description of protocols, like in state truncation \cite{Barnett1,Barnett2}. Let us start with an example in order to clarify this approach in a simple case: we consider a perfect photon counter able to discern the number of absorbed photons. If this detector displays the result \textit{n counts}, the checked proposition is the projector $\hat{P}_{n}=\vert n\rangle\langle n\vert$ on the photon-number state $\vert n\rangle$, and the pre-measurement state is precisely this photon-number state $\hat{\rho}_{ n}^{\mathrm{retr}}=\hat{P}_{n}$, with which we can make predictions about the preparations leading to such a result. Thus, for instance for preparations of the light in the photon-number states $\vert m\rangle$, the only possible preparation corresponds to $m=n$ for which the retrodicted probability is equal to 1.

According to the generalization of Gleason's theorem, the retrodictive approach also needs a state and propositions. The state is assigned to the system on the basis of the measurement result \textit{n}. This is the retrodicted state $\hat{\rho}^{\mathrm{retr}}_{n}$ with which we make retrodictions about state preparations leading to this result. The propositions about this state simply correspond to the different preparations of the measured system before its interaction with the apparatus. 
We note $\hat{\Theta}_{m}$ the proposition operator associated to each preparation choice \textit{m}, and in order to have an \textit{exhaustive set} of propositions, these operators $\hat{\Theta}_{m}$ should be a resolution of the Hilbert space, $\sum_{m}\,\hat{\Theta}_{m}=\hat{1}$.
Thus, the retrodictive probability of preparing the system in the state $\hat{\rho}_{m}$ when the measurement gives the result \textit{n} can be written as
\begin{equation}
\label{retrodictive_proba}
\mathrm{Pr}\left(m\vert n\right)=\mathrm{Tr}\lbrace\hat{\rho}^{\mathrm{retr}}_{n}\hat{\Theta}_{m}\rbrace.
\end{equation}

\textit{Bayesian Inference.$\--$} The expressions of the retrodicted states and proposition operators are given by a Bayesian inference from Born's rule which gives the predictive probabilities $\mathrm{Pr}\left(n\vert m\right)$ of having the result \textit{n} when we prepare the state $\hat{\rho}_{m}$, as previously derived by Barnett et al.\cite{Barnett1}. However, this derivation was based heretofore on an analogy with the form of Born's rule and not on the mathematical justification of the final form of retrodictive probabilities (\ref{retrodictive_proba}), given here by the Gleason's theorem. This theorem also justifies the strong requirement on the proposition operators $\hat{\Theta}_{m}$ to be an exhaustive set.
Thus, retrodiction requires no other additionnal assumption than the projection rule which may be used in state preparations. Indeed, Bayes' theorem leads for the retrodictive probability to the relation:
\begin{equation}
\label{retro_proba}
\mathrm{Pr}\left(\hat{\rho}_{m}\vert n\right) = \mathrm{Pr}\left(n\vert \hat{\rho}_{m}\right)\mathrm{Pr(\hat{\rho}_{m})}/\mathrm{Pr}(n),
\end{equation}
where the marginal probability $\mathrm{Pr}\left(n\right)$ of having the result 'n' is obtained by summing the joint probability on all 'candidate' states, 
\begin{equation}
\label{marginal_proba}
\mathrm{Pr}\left(n\right)=\sum_{m}\,\mathrm{Pr}\left(n\vert \hat{\rho}_{m}\right)\mathrm{Pr(\hat{\rho}_{m})}= \mathrm{Tr}\lbrace\hat{\rho}_{probe}\hat{\Pi}_{n} \rbrace,
\end{equation}
in which we define the state $\hat{\rho}_{probe}=\sum_{m}\,\mathrm{Pr}(m)\,\hat{\rho}_{m}$.
In a conditional preparation, such a state corresponds to an 'unread' measurement \cite{Haroche2006} : a mixture of states conditioned on each result 'm', weighted by their respective success probabilities $\mathrm{Pr}(m)$. 

Then, in order to write the retrodictive probability (\ref{retro_proba}) in its most general form (\ref{retrodictive_proba}) for a finite-dimensional Hilbert space $\mathcal{H}$, this statistical mixture should be such that $\hat{\rho}_{probe}=\hat{1}/D$, with $D=\mathrm{dim}\,\mathcal{H}$.
One can then identify the retrodicted state
\begin{equation}
\label{retro_state}
\hat{\rho}^{\mathrm{retr}}_{n}=\hat{\Pi}_{n}/\mathrm{Tr}\lbrace\hat{\Pi}_{n}\rbrace,
\end{equation}
and the proposition operators which are given by $\hat{\Theta}_{m}=D\mathrm{Pr}(m)\hat{\rho}_{m}$.

Finally, let us note that the retrodicted state can be propagated backward in time using the usual quantum evolution or propagation equations \cite{PRLBarnett, Barnett2}. This feature can be easily understood since the retrodicted state is obtained from an observable  (i.e. here a POVM) which evolves backward in time in the Heisenberg picture of the predictive approach. 

\textit{Quantum properties of a measurement revealed by the retrodictive approach.$\--$}
The retrodictive approach provides now a well-defined quantum state associated with a measurement outcome, on which usual quantum state analysis can be applied. The quantum properties of a measurement can be translated into properties of this state, as we see now.

\textit{1-Projectivity and Ideality of a measurement.}
We first introduce two estimators which are intrinsic to the apparatus. An 'ideal' measurement checks a simple proposition corresponding to a projector $\hat{\Pi}_{n}=\vert\psi_{n}\rangle\langle\psi_{n}\vert$ in the Hilbert space.
However, in more realistic situations, a measuring device is characterized by POVM elements which are not at all projectors.  An evaluation of \textit{the projectivity of a measurement} is given by the purity $\pi_{n}$ of its retrodicted state:
\begin{equation}
\label{projectivity}
\pi_{n}=\mathrm{Tr}\left[\left(\hat{\rho}^{\mathrm{retr}}_{n}\right)^{2}\right].
\end{equation}
When this state is pure, i.e. $\pi_{n}=1$, the measurement performed by the apparatus is  projective for the response n.

However, a projective measurement may be \textit{non-ideal}. The POVM element corresponding to a projective measurement with $\pi_{n}=1$ can be written as
\begin{equation}
\hat{\Pi}_{n}=\zeta_{n}\,\vert\psi_{n}\rangle\langle\psi_{n}\vert
\end{equation}
where $\zeta_{n}=\mathrm{Tr}\lbrace\hat{\Pi}_{n}\rbrace$.  By using the predictive probability $\mathrm{Pr}\left(n\vert\psi_{n}\right)=\zeta_{n}$, the parameter $\zeta_n$ can be interpreted as the detection efficiency of the retrodicted state, which is an intrinsic property of the apparatus. We evaluate the \textit{ideality of a measurement} by this detection efficiency.

In the more general case of a non-projective measuremement, $\zeta_n$ is given using (1) by
\begin{equation}
\zeta_{n}=\mathrm{Pr}\left(n\vert\hat{\rho}_{n}^{retr}\right)
=\mathrm{Tr}\lbrace\hat{\Pi}_{n}^{2}\rbrace/\mathrm{Tr}\lbrace\hat{\Pi}_{n}\rbrace=\pi_{n}\,\mathrm{Tr}\lbrace\hat{\Pi}_{n}\rbrace\le 1.
\end{equation}
This relation connects the projectivity and the ideality of a measurement.

It must be emphasized that the projective character of a measurement is revealed within the retrodictive approach, and not by the predictive one, for which the usual definition \cite{Neumann1955} of a projective measurement is the certainty of the result of two successive projective measurements. Such a measurement corresponds to a very particular case: a projective and ideal measurement, characterized by $\pi_{n}=\zeta_{n}=1$.

These two estimators provide a way to categorize the different kinds of detectors. The first category corresponds to apparatus performing a far from projective measurement. For instance, an avalanche photodiode is unable to discriminate between one or more photons \cite{Robert}.  A second category is when the measurement is highly projective and close to ideal. A photon counter able to discrimate the numbers of photons, and detect them with high efficiency, would correspond to this case. Finally, as our study highlights, it exists a third category corresponding to a projective but non-ideal measurement. In this case, the retrodicted state is  close to a pure state but its detection efficiency is less than unity. In this case, the result of two succesive projective, but non-ideal, measurements becomes uncertain.  For instance, conventional photon counters do not enter into this category, which includes more 'exotic' devices \cite{AmriPhD}. 

\textit{2-Fidelity with a projective measurement and Detectivity of a target state.}
After the definition of two intrinsic estimators, we now introduce quantities which depend on a given state $\vert\psi_{\mathrm{tar}}\rangle$, called in the following target state. We first define here the \textit{fidelity with a projective measurement on the target state} as the overlap between the state $\hat{\rho}^{\mathrm{retr}}_{n}$ retrodicted from a certain result \textit{n} and the target state, in which we would like checking the system before its interaction with the apparatus.
Such a fidelity can be written as \cite{Jozsa94}
\begin{equation}
\label{fidelity}
\mathcal{F}_{n}\left(\psi_{\mathrm{tar}}\right)=\langle\psi_{\mathrm{tar}}\vert\hat{\rho}^{\mathrm{retr}}_{n}\vert\psi_{\mathrm{tar}}\rangle.
\end{equation}
The retrodictive approach provides an interesting interpretation for this overlap. This is the retrodictive probability of preparing the system in the target state $\vert\psi_{\mathrm{tar}}\rangle$, before the measurement process giving the result \textit{n}:
\begin{equation}
\label{fidelity_proba}
\mathcal{F}_{n}\left(\psi_{\mathrm{tar}}\right)=\mathrm{Pr}\left(\psi_{\mathrm{tar}}\vert n\right)=\mathrm{Tr}\lbrace\hat{\rho}^{\mathrm{retr}}_{n}\hat{\Theta}_{\mathrm{tar}}\rbrace.
\end{equation}
The proposition operator about the state of the system, just after its preparation, is $\hat{\Theta}_{\mathrm{tar}}=\vert\psi_{\mathrm{tar}}\rangle\langle\psi_{\mathrm{tar}}\vert$.
When the measurement giving the result \textit{n} is sufficiently faithful $\mathcal{F}_{n}\left(\psi_{\mathrm{tar}}\right)\simeq 1$, the most probable state in which the system was prepared before its interaction with the apparatus is this target state $\vert\psi_{\mathrm{tar}}\rangle$.

On the other hand, we introduce what we call the \textit{detectivity of the target state}, given by the predictive probability of detecting it. using (1) and the expression (5) of the retrodicted state, it can be written as:
\begin{equation}
\kappa_n=\mathrm{Pr}\left(n|\psi_{\mathrm{tar}}\right)=\mathrm{Tr}\lbrace\hat{\Pi}_n\rbrace\,\mathrm{Tr}\lbrace\hat{\rho}_{\mathrm{tar}}\,\hat{\rho}^{\mathrm{retr}}_{n}\rbrace.
\end{equation}
This estimator corresponds to the detection efficiency $\zeta_n$ if the state to detect is the retrodicted state. For single-photon counters for instance, the detectivity of a single-photon state corresponds to the usual definition of 'quantum efficiency'. 

Our study enables to connect these quantities with the intrinsic estimators defined before. By using (8), we find an interesting relation between all the estimators: 
\begin{equation}
\kappa_n=\zeta_n\, \mathcal{F}_n/\pi_n.
\end{equation}
It can be seen from this expression that, for an ideal measurement, the fidelity can only enhance the detectivity of the target state. However, for a given ideality and fidelity, a less projective measurement may be better for detecting it. Indeed, the retrodicted state is then mixed and the measurement becomes less discriminating.

\textit{3-Non-classicality of a measurement.}
There are different manifestations for the non-classicality of states. For instance in quantum optics, this property is revealed by different signatures \cite{Treps}: variances below the standard quantum noise, upper bound for eigenvalues of the covariance matrix \cite{Simon}, negativity in particular quasi-probability distributions \cite{tomography}. Whatever the signature that is used, the non-classicality of a measurement corresponds to the non-classicality of the retodicted state.

We illustrate the relevance of such a correspondance for optical detectors in the conditional preparation of non-classical states of light \cite{Lvovsky2001,laurat2003, Ourjoumtsev2006}.
For such protocols, two light beams A and B in an entangled state $\hat{\rho}_{AB}$ are necessary, since the involved measurements are generally destructive. When we perform a judicious measurement on the beam B, the state of the other beam, conditionned on an expected result "n", is given by the projection rule
\begin{equation}
\label{cond_state}
\hat{\rho}_{A,n}^{\mathrm{cond}}
=\frac{1}{\mathrm{Pr}\left(n\right)}\mathrm{Tr}_{B}\lbrace\hat{\rho}_{AB}\,\hat{1}_{A}\otimes\hat{\Pi}_{n}\rbrace
\end{equation}
where $\hat{\Pi}_{n}$ is the POVM element describing the measurement given the result "n". In quantum optics, the gaussian entanglement si a very universal resource. For instance, the two-mode state generated by spontaneous parametric down-conversion, can be written on the photon-number states as 
\begin{equation}
\label{EPR_state}
\vert\psi_{\mathrm{AB}}\rangle = \left(1-\lambda^{2}\right)^{1/2}\sum_{n=0}^{\infty}\,\lambda^{n}\vert n,n\rangle,
\end{equation}
where $\lambda$ varies between 0 and 1. If the limit of high-gain, i.e. $\lambda\simeq 1$, the conditional state becomes:
\begin{equation}
\label{cond_state_gaussian}
\hat{\rho}_{\mathrm{A, cond}}^{[n]}\underset{\lambda\rightarrow 1}\simeq \frac{\hat{\Pi}_{n}^{*}}{\mathrm{Tr}\lbrace\hat{\Pi}_{n}\rbrace}=\left(\hat{\rho}_{n}^{retr}\right)^{*}.
\end{equation}
We recognize the complex conjugate of the state retrodicted from the result 'n'. Thus, a necessary condition for preparing more exotic non-classical states is the use of measurement devices characterized by non-classical retrodicted states. 

An interesting link can also be established between the Gaussian character of a measurement and its projectivity. The Hudson-Piquet's theorem \cite{Hudson1974} states that any pure state characterized by a non-negative Wigner representation has a Gaussian Wigner representation. We deduce from it that, when a measurement is projective and is characterized by a retrodictive state having a non-negative Wigner representation, then the measurement is Gaussian. 

\textit{Conclusion.$\--$} In this paper, we have shown that the retrodictive approach of quantum physics provides an interesting way to characterize the quantum properties of a measurement apparatus: these properties are obtained from the ones of the retrodicted state. This approach, by introducing a well-defined quantum state and making use of traditional quantum state analysis, allows to define precise estimators for qualifying a measurement. It introduces subtle distinctions, such as projectivity and ideality, which were more qualitative heretofore. We provide here a formal definition of these properties and new insights on them thanks to the retrodictive approach. Such definition should lead to further analysis of quantum measurement apparatus and facilitate the design of detectors able to detect specific quantum states, with potential applications in metrology for instance.


\begin{thebibliography}{99}
\bibitem{Neumann1955} J. Von Neumann, \textit{Mathematical foundations of quantum mechanics}, Princeton University Press (1955).
\bibitem{Zurek1982} J. A. Wheeler and W. H. Zurek, \textit{Quantum Theory of measurement}. Princeton University Press, NJ (1982).
\bibitem{KLM} E. Knill, R. Laflamme, G.J. Millburn, Nature \textbf{409}, 46-52 (2001).
\bibitem{OBrien} J.L. O'Brien, Science \textbf{318}, 1567 (2007).
\bibitem{mandel} C.K. Hong and L. Mandel, Phys. Rev. Lett. \textbf{56}, 58 (1986).
\bibitem{Ourjoumtsev2007} A. Ourjoumtsev et al., Science \textbf{312}, 83 (2006).
\bibitem{QDT1} J. Lundeen et al., Nature Physics \textbf{5}, 27 (2009).
\bibitem{Helstrom} C.W. Helstrom, \textit{Quantum detection and estimation theory}, Academic Press, NY (1976).
\bibitem{QDT2} J. Fiurasek, Phys. Rev. A \textbf{64}, 024102 (2001).
\bibitem{Aharonov} Y. Aharonov, P. G. Bergman, J. L. Lebowitz, Phys. Rev. \textbf{134}, B1410 (1964).
\bibitem{Barnett0} S. M. Barnett et al., J. Mod. Opt. \textbf{47}, 1779 (2000).

\bibitem{Busch} P. Busch, Phys. Rev. Lett. \textbf{91}, 120403 (2003).
\bibitem{Gleason1957} A.M. Gleason, J. Math. Mech. \textbf{6}, 885 (1957).
\bibitem{Barnett1} D.T. Pegg and S.M. Barnett, J. Opt. B: Quantum Semiclass. Opt  \textbf{1}, 442 (1999).
\bibitem{Barnett2} S. M. Barnett and D. T. Pegg, Phys. Rev. A \textbf{60}, 4965 (1999).
\bibitem{Haroche2006} S. Haroche and J.M. Raimond, \textit{Exploring the quantum: Atoms, Cavities, and Photons}, Oxford Graduate Texts (2006).
\bibitem{PRLBarnett} S.M. Barnett et al., Phys. Rev. Lett. \textbf{86}, 2455 (2001).

\bibitem{Robert} R.H. Hadfield, Nature Photon. \textbf{3}, 696 (2009).

\bibitem{AmriPhD} T. Amri, PhD Thesis, to be published


\bibitem{Jozsa94} R. Jozsa, J. Mod. Opt. \textbf{41}, 2315 (1994). 

\bibitem{Treps} N. Treps and C. Fabre, Laser Physics \textbf{15}, 187 (2005).
\bibitem{Simon} B. Arvind, B. Dutta, N. Mukunda, R. Simon, Pramana J. Phys. \textbf{45}, 471 (1995),  eprint arXiv :quant-ph0701221v2 (2007).
\bibitem{tomography} U. Leonhardt, \textit{Measuring the quantum state of light}, Cambridge University Press, Cambridge (1997).

\bibitem{Lvovsky2001} A.I. Lvovsky et al., Phys. Rev. Lett. \textbf{87}, 050402 (2001).
\bibitem{laurat2003} J. Laurat et al., Phys. Rev. Lett. \textbf{91}, 213601 (2003).

\bibitem{Ourjoumtsev2006} A. Ourjoumtsev et al., Phys. Rev. Lett. \textbf{96}, 213601 (2006).
\bibitem{Hudson1974} R. L. Hudson, Rep. Math. Phys. \text{6}, 249 (1974).

\end{thebibliography}
\end{document}